\documentclass{epl}
\usepackage{psfig}

\title{Disorder-driven phase transitions of the large $q$-state Potts model in three dimensions}
\shorttitle{3$d$ random bond Potts model}
\author{M. T. Mercaldo\inst{1} \and J-Ch. Angl\`es d'Auriac\inst{2} \and
F. Igl\'oi\inst{3,4}}
\institute{
  \inst{1} Dipartimento di Fisica ``E.R. Caianiello'', Universit\`a degli Studi di
Salerno and  Laboratorio Regionale SuperMat INFM, I-84081 Baronissi (Salerno), Italy\\
  \inst{2} Centre de Recherches sur les Tr\'es Basses
Temp\'eratures,
B. P. 166, F-38042 Grenoble,
France\\
  \inst{3} Research Institute for Solid State Physics and Optics,
H-1525 Budapest, P.O.Box 49, Hungary\\
  \inst{4} Institute of Theoretical Physics,
Szeged University, H-6720 Szeged, Hungary
}

\begin{document}

\maketitle

\begin{abstract}
  Phase transitions induced by varying the strength of disorder in the
  large-$q$ state Potts model in 3d are studied by analytical and
  numerical methods. By switching on the disorder the transition stays
  of first order, but different thermodynamical quantities display
  essential singularities. Only for strong enough disorder the
  transition will be soften into a second-order one, in which case the
  ordered phase becomes non-homogeneous at large scales, while the
  non-correlated sites percolate the sample. In the critical regime
  the critical exponents are found universal: $\beta/\nu=0.60(2)$ and
  $\nu=0.73(1)$.
\end{abstract}

\newcommand{\bc}{\begin{center}}
\newcommand{\ec}{\end{center}}
\newcommand{\be}{\begin{equation}}
\newcommand{\ee}{\end{equation}}
\newcommand{\beqn}{\begin{eqnarray}}
\newcommand{\eeqn}{\end{eqnarray}}

Real materials are generally three-dimensional and contain some sort
of quenched disorder, whose effect on the properties of the
phase transition is an intensively studied question. If the phase
transition in the non-random system is of second order, relevance or
irrelevance of disorder can be decided due to the Harris
criterion~\cite{harris} and also perturbative field-theoretical
calculations are possible~\cite{perturbative}. For first-order
transitions in the pure system, however, we have only a more limited
knowledge. In 2d any amount of continuous disorder is
sufficient to soften the transition into a second-order one~\cite{aizenmanwehr,imrywortis},
which is studied in detail numerically~\cite{pottsmc,pottstm,long2d}.
In 3d, however, there is no such a rigorous
criterium. Here experimental work~\cite{exp} and numerical
simulations~\cite{uzelac,pottssite,pottsbond} show that for strong
enough disorder the phase transition becomes second order. In the
numerical work mostly the $q=3-$ and $4$-state Potts models were
studied by Monte-Carlo simulations, both by
site~\cite{uzelac,pottssite} and bond dilutions~\cite{pottsbond}.

In the present work we revisit the problem of the effect of disorder
at first-order phase transitions in 3d and address mainly such questions
which have not been treated previously. We
consider the disorder as realized by random ferromagnetic bonds, which
contains as a limiting case the bond dilution. The
problem we investigate is the Potts model in the large-$q$ limit, which allows us
to make partially analytical calculations and in this way to shed
light on the physical mechanism responsible of the softening of
the transition. We have also performed extensive numerical calculations,
in which exact results on large finite samples are averaged over disorder.
In particular, we addressed the question of universality of the second order transition
and the location of the tricritical point.

We start to define the Hamiltonian of the $q$-state Potts model~\cite{Wu}:
\begin{equation}
\mathcal{H}=-\sum_{\left\langle i,j\right\rangle }J_{ij}\delta(\sigma_{i},\sigma_{j})
\label{eq:hamilton}
\end{equation}
in terms of the Potts-spin variables, $\sigma_{i}=0,1,\cdots,q-1$, at
site $i$ and the summation runs over nearest neighbors of a simple
cubic lattice.  The ferromagnetic couplings, $J_{ij}>0$, can take two
values, $J_1=J(1+\delta)$ and $J_2=J(1-\delta)$ with equal
probability~\cite{discrete}. The strength of disorder is $\delta$ and
the dilute model is recovered for $\delta=1$.  As usual for the random
model we calculate disorder averages of the free energy and the
correlation function.

For a given realization of disorder the partition function is given in the random cluster
representation~\cite{kasteleyn}:
\begin{equation}
Z =\sum_{G}q^{c(G)}\prod_{ij\in G}\left[q^{\beta J_{ij}}-1\right]
\label{eq:kasfor}
\end{equation}
where the sum runs over all subset of bonds, $G$, $c(G)$ stands for
the number of connected components of $G$ and
we use the reduced temperature~\cite{long2d}, $T \to T \ln q=O(1)$ and its inverse as $\beta
\to \beta/\ln q$.
In the large-$q$ limit, where $q^{\beta J_{ij}} \gg 1$, the partition function can be written as
\begin{equation}
Z=\sum_{G\subseteq E}q^{\phi(G)},\quad \phi(G)=c(G) + \beta\sum_{ij\in G} J_{ij}\label{eq:kasfor1}
\end{equation}
which is dominated by the largest term, $\phi^*=\max_G \phi(G)$, and
the degeneracy of the optimal set $G^*$ is likely to be one. Thus the study of
the large-$q$ state Potts model is reduced to an optimization problem, whose solution
(for a given realization of disorder) depends on the temperature. The
free-energy per site is proportional to $\phi^*$ and given by $-\beta
f= \phi^*/N$, where $N$ stands for the number of sites of the
lattice. From the point of view of magnetization and the correlation
function the largest connected cluster of the optimal set, $\cal{C}$,
plays a crucial role. The magnetization, $m$, is given by the fraction
of sites in $\cal{C}$, and its disorder average is $m>0$ in the ordered phase, $T < T_c$. In
the disordered phase, $T > T_c$, $\cal{C}$ has a finite linear extent,
$\xi < \infty$, which is proportional to the average correlation length in the
system. At the phase-transition point, $T = T_c$, if the transition is
of second order, $\cal{C}$ is a fractal and its fractal dimension,
$d_f$, is related to the magnetization scaling dimension, $x_m$, as
$d=d_f+x_m$. Here $x_m$ can be expressed by the standard critical
exponents as $x_m=\beta/\nu$. On the contrary, if the transition is of
first order there is phase coexistence at $T = T_c$ and the
correlation length is finite. This type of formalism has been used by
us previously in 2d~\cite{long2d}, and we refer to this work for a detailed presentation
of the method.

The solution of the problem is the simplest for the non-random model,
$\delta=0$, when there are only two homogeneous optimal sets. For
$T<T_c(0)$ it is the fully connected diagram with a free-energy:
$-\beta N f=1+N\beta J z$ ($z=d=3$ is the connectivity of the lattice)
and for $T>T_c(0)$ it is the empty diagram with $-\beta N
f=N$. Consequently the transition point is located at:
$T_c(0)=Jz/(1-1/N)$ and the latent heat is $\Delta e/T_c(0)=1$.

Introducing disorder into the system, $\delta>0$, new type of
non-homogeneous optimal diagrams will appear close to $T_c(0)$. Let us
consider first $T>T_c(0)$, where the new optimal set contains
connected clusters of linear size, $l$, embedded into the empty
diagram. Here we refer to the treatment of this problem in
2d~\cite{long2d} when close to $T=T_c(0)$ the excess quantities over
the homogeneous optimal sets are i) a bulk term due to disorder
fluctuations, $\Delta f_b$, and ii) an interface term due to missing
bonds, $\Delta f_s \sim \beta JS$, where the size of the interface is $S \sim
l^{d-1}$. For Gaussian fluctuations $\Delta f_b^G \sim \delta \beta J
V^{1/2}$, where $V \sim l^d$ is the volume of the cluster.
Consequently for $d \le 2$ normal fluctuations of disorder will lead
to the creation of clusters of arbitrary size and for any weak
disorder there is no phase coexistence and the transition is of second
order~\cite{aizenmanwehr}.

On the contrary for $d>2$, clusters are created only due to extreme
fluctuations, when say the cluster consists of only strong bonds,
thus  $\Delta f_b^+ \sim \beta J\delta (V-cS)$, where $c=O(1)$ is a geometrical
factor. The existence of such large cluster is exponentially rare,
its density is given by $-\ln \rho_{+} \sim V-cS$. For a small $\delta$
only large clusters can be formed and  we obtain the possible size
from the condition, $\Delta f_b^+ \ge \Delta f_s$, as $l \ge l_+(\delta) \sim 1/\delta$\cite{note1}.
Finally, the free energy of the non-homogeneous optimal set relative to the empty graph is given by
the sum of contribution of the possible clusters:
\be
-\beta N f_+ \simeq N+ N\sum_{l \ge l_+(\delta)} (\Delta f_b^+(l) - \Delta f_s(l))\rho_{+}(l)\;,
\label{f+}
\ee
and it is dominated by the contribution with $l=l_+(\delta)$.

A similar analysis can be performed for $T<T_c(0)$, when the new inhomogeneous optimal set
is obtained from the fully connected graph by creating clusters of isolated sites. In this case
a cluster of size, $l$, has $V+cS$ weak bonds, a bulk gain of $\Delta f_b^- \sim \beta J\delta (V+cS)$
and appears with a density, $-\ln \rho_{-} \sim V+cS$.
Note that there is a difference in the sign of the surface term comparing with the high-temperature
case. The limiting size of clusters, $l_-(\delta)$
can be estimated as above and the free energy of the optimal set can be written in an analogous
form with eq.~(\ref{f+}):
\be
-\beta N f_- \simeq 1+N\beta Jz+N \sum_{l \ge l_-(\delta)} (\Delta f_b^-(l) - \Delta f_s(l))\rho_{-}(l)\;,
\label{f-}
\ee
and it is dominated by the contribution with $l=l_-(\delta)$. The asymmetry in the densities in the two
phases, $\rho_{+}$ and $\rho_{-}$, leads to a shift in the critical temperature,
$\ln (T_c(\delta)-T_c(0)) \sim -\delta^{-3}$,
and similarly to a reduction of the latent heat and the jump of the
magnetization at the transition point:
\begin{equation}
\ln (1-\Delta e/T_c) \sim \ln(1- \Delta m) \sim-\frac{1}{\delta^3}\;.
\end{equation}
Thus the phase transition stays first order and as the disorder is
switched on there is an essential singularity in the thermodynamical
quantities as a function of $\delta^{-3}$.

For finite values of disorder the optimal set (and thus the free energy and
the magnetization) is exactly calculated by a
combinatorial optimization algorithm~\cite{aips02}, which works in strongly polynomial time.
In the numerical calculations we studied periodic systems of cubic shape with a linear
size $L=16,24,32$ and in some cases $L=40$. For the largest size the computation of the optimal
set for a single sample
typically took from $5$ to $20$ hour CPU time in a 2.8GHz processor. (Since the breaking-up length~\cite{note1}
at $\delta=0.40$ is $L_{br} \approx 40$ we restricted ourselves to $0.40 \le \delta \le 1$.)
The number of realizations were several thousands, for the
largest size several hundreds.

\begin{figure}
 \onefigure[width=8cm,angle=270]{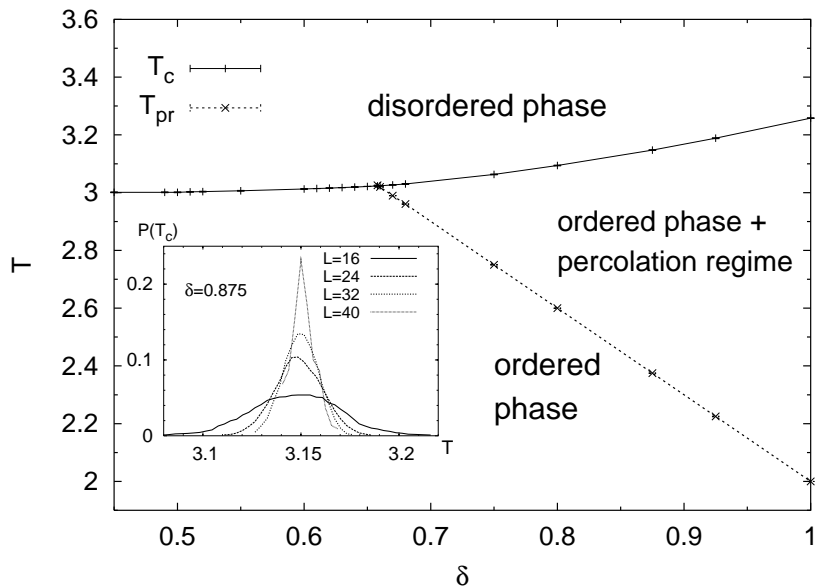}
 \caption{Phase diagram of the random-bond Potts model for bimodal disorder. The calculated
transition points are denoted by crosses the size of which is larger than the error of the
calculation. In the ordered phase
     in the optimal
     set the non-connected sites are percolating for $\delta>\delta_{pr}$
     and $T_{pr}(\delta)<T<T_c(\delta)$, see eq.~(\ref{T_pr}).
     Numerical results in fig. \ref{magn} indicate, that
     the first- and second-order transition regimes are separated by a
     critical disorder, $\delta_{c}$, which corresponds to the
     border of the percolation regime, $\delta_{c}=\delta_{pr}$. Inset: Distribution of
     the finite-size transition points identified with the temperature at which the
     connected cluster starts to percolate.}
 \label{phase-d}
\end{figure}

We start to show in fig. \ref{phase-d} the numerically calculated
phase diagram as a function of the strength of disorder. Here the
phase boundary between the ordered and disordered phases is obtained
from the condition that the largest cluster in the optimal set starts
to percolate and the finite-size results are then extrapolated, see
the inset of fig. \ref{phase-d}. The
transition points obtained by this method agree, within the error of
the calculation, with those calculated from the position of the maxima
of the specific heat. Before analyzing the properties of the phase
transition we mention that the structure of the optimal set in the
ordered phase can be of two different types. For weak disorder,
$\delta < \delta_{pr}$, or for low temperature, $T < T_{pr}$, in the
optimal set the isolated sites form finite clusters, otherwise the
isolated sites are percolating. In the dilute model with $\delta=1$
and $J_1=2J$ the percolation temperature is at $T_{pr}(1)=J_1$. Indeed
the optimal set for $T<J_1$ contains all the strong bonds, whereas
from this optimal set for $T>J_1$ the dangling bonds are removed.
Since the dangling bonds represent a finite fraction of the bonds the
non-occupied sites become percolating.

For $\delta<1$ the skeleton of dangling bonds of $J_1$ couplings are decorated by weak $J_2$
couplings, and by removing one strong dangling bond one has also to remove four weak decorated
ones, which is possible in the temperature range:
\be
T_c>T>T_{pr}(\delta)=J(5-3 \delta)\;.
\label{T_pr}
\ee
The numerical results indicate that at $T_{pr}(\delta)$ in a finite fraction of samples
there is a giant cluster of isolated sites which spans the finite cube.

\begin{figure}
 \onefigure[width=8cm,angle=270]{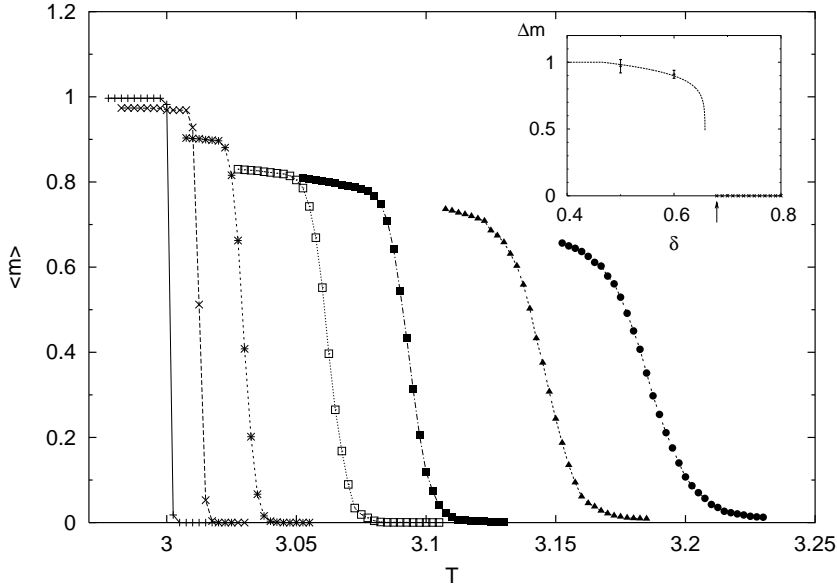}
 \caption{Temperature dependence of the magnetization for $L=16$ with different strength
of disorder: $\delta=0.5,~0.6,~0.68,~0.75,~0.8,~0.875,~0.925$ from left to right. Inset: Jump of
the magnetization at the transition point as extrapolated at $\delta=0.4,~0.5$ and $0.6$.
In the region to the right of the arrow, $\delta \ge 0.68$, the fractal dimension
is obtained $d_f<3$ (see the method in fig.\ref{scaling})
which indicates a second order transition, thus $\Delta m=0$, which is shown by crosses. The
line drawn over the extrapolated points is a guide to the eyes.
}
 \label{magn}
\end{figure}

The temperature dependence of the magnetization for different values of the disorder is
illustrated in fig. \ref{magn} in a finite system of $L=16$. For $\delta < \delta_c$ in
the first-order transition regime the magnetization has a finite jump, $\Delta m>0$.  $\Delta m(L)$
is estimated from the relation: $\Delta m(L)\approx 2 m[T_c(L)]$, and the
extrapolated values are shown in the inset of fig. \ref{magn} for different values
of $\delta$. From this we estimate the position of the tricritical disorder
$\delta_c=0.65(2)$. It is interesting to observe that
this value, within the relatively large error coincides with the limiting value of disorder,
$\delta_{pr}=0.658(1)$,
where the percolation of isolated sites in the ordered phase starts. Physically it is very
intuitive that percolation of isolated sites is the prerequisite of a continuous transition
in the system. Indeed
the correlation length in the ordered phase, which is measured by the linear size of the largest
finite (i.e. not the giant) ordered cluster can be divergent, if it is embedded in an infinite
cluster of isolated sites. Therefore we tend to conjecture that $\delta_c=\delta_{pr}$.

\begin{figure}
  \onefigure[width=8cm,angle=270]{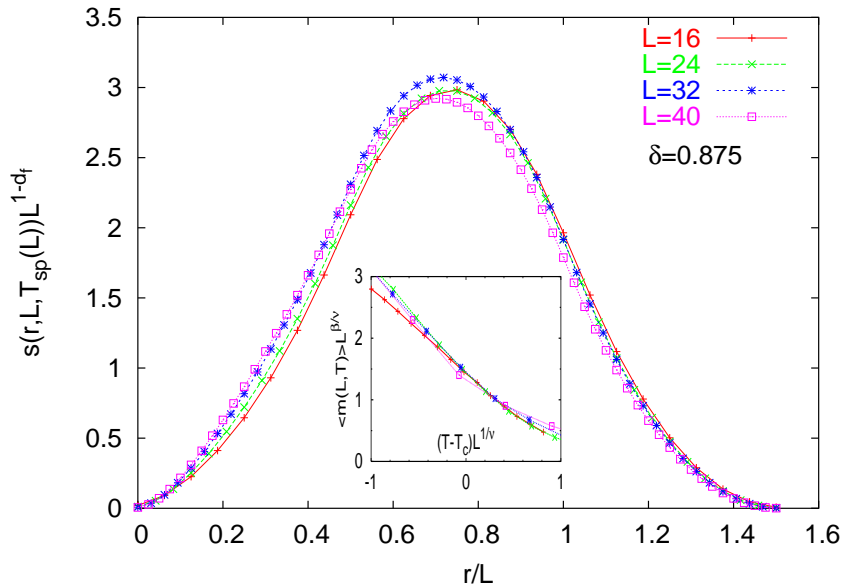}
  \caption{Scaling plot of the mass of a shell of the giant cluster, for each size at the average
spanning temperature, see text. Optimal collapse is obtained with a fractal dimension, $d_f=2.40$.
Inset: Scaling plot of the magnetization as a function of the distance from the
critical temperature. By fixing $\beta/\nu=d-d_f=0.60$ the best collapse is obtained with $\nu=0.73(1)$}
  \label{scaling}
\end{figure}

In the second-order transition regime we have made the most detailed calculations at $\delta=7/8$,
but universality is checked at other values of the disorder at $\delta=0.75,~0.80,~0.925$ and $1$.
The fractal dimension, $d_f$, of the percolating connected cluster is estimated in such a way
that fixing a reference point we measured the average number of points, $s(r,L,T)$, within
a shell around the reference point of unit width and radius, $r$. Close to the transition
point: $t=(T-T_c)/T_c \ll 1$ it is expected to scale as:
\begin{equation}
s(r,L,t)=L^{d_f-1} \tilde{s}(r/L,tL^{1/\nu})\;.
\end{equation}
Since the exact value of the critical temperature is not known we can not fix the second
argument of the scaling function as $\tilde{s}(\rho,\tau=0)$. Instead
we set $T=T_{sp}(L)$, which is the temperature, where the connected cluster spans the cube
of the given size and we average over disorder. Evidently with this choice the second argument
of the scaling function, $\tau$, is asymptotically constant and the
scaling function depends only on one parameter: $\tilde{s}=\tilde{s}(r/L)$. Our scaling
picture is verified in fig. \ref{scaling} in which
the scaling plot of the mass in the shell is shown. From the  optimal scaling collapse\cite{collapse}
we obtained for the fractal dimension, $d_f=2.40(2)$.

The correlation length critical exponent, $\nu$, is calculated from the scaling behavior of
the magnetization: $m(t,L)=L^{x_m} \tilde{m}(tL^{1/\nu})$.
From an optimal scaling collapse\cite{collapse} as
shown in the inset of fig. \ref{scaling} we have obtained $\nu=0.73(1)$.
Note that $\nu$ satisfies the rigorous bound for disordered systems~\cite{ccfs}: $\nu \ge 2/d$.
Due to strong cross-over effects we could not determine the tricritical exponents with
sufficient accuracy. However, the tricritical correlation length exponent,
$\nu_{tr}$, is related\cite{pottstm} to the exponents of the random-field Ising model as:
$\nu_{tr}=\nu_{RF}/(2-\alpha_{RF}-\beta_{RF})$. Using numerical estimates\cite{RFIM} we
obtain $\nu_{tr} \approx 0.69$.

In Table \ref{table:1} we have collected the presently known numerical results about the
critical exponents of random $q$-state Potts models in 3d, including also the Ising model ($q=2$)
and percolation, which formally corresponds to $q=1$. The $q$-dependence of the
exponents, in particular that of $\beta/\nu$ is non-monotonic.

\begin{table}
\caption{Critical exponents of the random $q$-state Potts model in 3d comparing with
those of percolation}
\label{table:1}
\begin{center}
 \begin{tabular}{|c|c|c|c|c|c|}  \hline
   & perc.\cite{staufferaharony}& $q=2$~\cite{ising3d}& $q=3$~\cite{pottssite}
 & $q=4$~\cite{pottsbond} & $q \to \infty $ \\ \hline
  $\nu$ & $0.877$ & $0.684(5)$ & $0.690(5)$ & $0.747(10)$ & $0.73(1)$  \\
  $\beta/\nu$ & $0.477$ & $0.518(2)$ & $0.539(2)$ & $0.732(20)$ & $0.60(2)$ \\ \hline
  \end{tabular}
 \end{center}
\end{table}

In our
model thermal fluctuations play a negligible role and the physical
properties are solely determined by disorder effects, which
 is characteristic at a so called infinite
disorder fixed point~\cite{reviewcf}. In the two-dimensional
problem an isomorphism is conjectured~\cite{long2d} with the phase transition in the strongly
anisotropic random transverse-field Ising model in 1d. However, our numerical results indicate that
this type of relation does not seem to work in higher dimensions.

\acknowledgments
This work has been supported by the French-Hungarian cooperation programme Balaton (Minist\`ere des
Affaires Etrang\`eres - OM), the Hungarian National
Research Fund under  grant No OTKA TO34183, TO37323, TO48721,
MO45596 and M36803, F.I. thanks the Dipartimento di Fisica ``E.R. Caianiello''
Universit\`a degli Studi di Salerno for hospitality.


\begin{thebibliography}{0}

\bibitem{harris}
  \Name{Harris A.~B.}
  \REVIEW{J. Phys. C}{7}{1974}{1671}.

\bibitem{perturbative}
 \Name{Ludwig A.~W.~W.}
 \REVIEW{Nucl. Phys. B}{285}{1987}{97},
 \SAME{330}{1990}{639};
 \Name{Ludwig A.~W.~W. \and Cardy J.}
 \SAME{285}{1987}{687};
 \Name{Dotsenko Vl., Picco M. \and Pujol P.}
 \REVIEW{Nucl. Phys. B}{455}{1995}{701}.

\bibitem{aizenmanwehr}
  \Name{Aizenman M. \and Wehr J.}
  \REVIEW{Phys. Rev. Lett.}{62}{1989}{2503}; {\it errata} \SAME{64}{1990}{1311}.

\bibitem{imrywortis}
  \Name{Imry Y. \and Wortis M.}
  \REVIEW{Phys. Rev. B}{19}{1979}{3580};
  \Name{Hui K. \and Berker A.~N.}
  \REVIEW{Phys, Rev. Lett.}{62}{1989}{2507}.

\bibitem{pottsmc}
  \Name{Picco M.}
  \REVIEW{Phys. Rev. Lett.}{79}{1997}{2998};
  \Name{Chatelain C. \and Berche B.}
  \REVIEW{Phys. Rev. Lett.}{80}{1998}{1670},
  \REVIEW{Phys. Rev. E}{58}{1998}{R6899},
  \SAME{60}{1999}{3853};
  \Name{Olson T. \and Young A.~P.}
  \REVIEW{Phys. Rev. B}{60}{1999}{3428}.

\bibitem{pottstm}
  \Name{Cardy J. \and Jacobsen J.~L.}
  \REVIEW{Phys. Rev. Lett.}{79}{1997}{4063};
  \Name{Jacobsen J.~L. \and Cardy J.}
  \REVIEW{Nucl. Phys. B}{515}{1998}{701}.

\bibitem{long2d}
  \Name{Angl\`es d'Auriac J.-Ch. \and  Igl\'oi F.}
  \REVIEW{Phys. Rev. Lett.}{90}{2003}{190601};
  \Name{Mercaldo M.~T., Angl\`es d'Auriac J.-Ch \and Igl\'oi F.}
  \REVIEW{Phys. Rev. E}{69}{2004}{056112}.

\bibitem{exp}
  \Name{Iannacchione G.~S., Crawford G.~P., Zumer S., Doane J.~W. \and Finotello D.}
  \REVIEW{Phys. Rev. Lett.}{71}{1993}{2595}.

\bibitem{uzelac}
  \Name{Uzelac K., Hasmy A. \and Jullien R.}
  \REVIEW{Phys. Rev. Lett.}{74}{1995}{422}.

\bibitem{pottssite}
  \Name{Ballesteros H.~G., Fern\'andez L.~A., Mart\`in-Mayor V., Mu\~noz Sudupe A.,
  Parisi G. \and Ruiz-Lorenzo J.~J.}
  \REVIEW{Phys. Rev. B}{61}{2000}{3215}.

\bibitem{pottsbond}
  \Name{Chatelain C., Berche B., Janke W. \and Berche P.~E.}
  \REVIEW{ Phys. Rev. E}{64}{2001}{036120};
  preprint cond-mat/0501115.

\bibitem{Wu}
  \Name{Wu F.~Y}
  \REVIEW{Rev. Mod. Phys}{54}{1982}{235}.

\bibitem{discrete}
The descrete form of the disorder could result in extra, non-physical singularities,
see in detail for the 2d case in~\cite{long2d}, which however does not influence
the properties of the phase transition.

\bibitem{kasteleyn}
  \Name{Kasteleyn P.~W.\and Fortuin C.~M}
  \REVIEW{J. Phys. Soc. Jpn.}{26}{1969}{(suppl.)11}.

\bibitem{aips02}
  \Name{Angl\`es d'Auriac J.-Ch, Igl\'oi F., Preissmann M. \and Seb\"o A.}
  \REVIEW{J. Phys. A}{35}{2002}{6973};
  \Name{Angl\`es d'Auriac J.-Ch}
  \Book{New Optimization Algorithms in Physics}
  \Editor{Hartmann A.~K. \and Rieger H.}
  \Publ{Wiley-VCH, Berlin}
  \Year{2004}
  \Page{101}.

\bibitem{note1}
  Note that there is a breaking-up length, $L_{br}(\delta)$, so that in a finite system with,
$L<L_{br}(\delta)$ the optimal set is homogeneous. This length can be estimated as
$[L_{br}(\delta)]^d \rho_+(l_+(\delta))=O(1)$.


\bibitem{collapse}
  Optimal scaling collapse is obtained by minimizing the area of the collapse region, see in
Ref\cite{long2d}, first paper fig. 2.

\bibitem{ccfs}
  \Name{Chayes J.~T., Chayes L., Fisher D.~S. \and Spencer T.}
  \REVIEW{Phys. Rev. Lett.}{57}{1986}{2999}.

\bibitem{RFIM}
  \Name{Rieger H. \and Young A.~P.}
  \REVIEW{J. Phys. A}{26}{1993}{5279};
  \Name{Rieger H.}
  \REVIEW{Phys. Rev. B}{52}{1995}{5659}.

\bibitem{reviewcf}
  \Name{Igl\'oi F. \and Monthus C.}
  \Review{Phys. Rep.}\Page{(to be published)}.

\bibitem{staufferaharony}
  See in \Name{D. Stauffer and A. Aharony} \Book{Introduction to Percolation Theory}
  \Publ{Taylor and Francis, London} \Year{1992}.

\bibitem{ising3d}
  \Name{Ballesteros H.~G., Fern\'andez L.~A., Mart\`in-Mayor V., Mu\~noz Sudupe A.,
  Parisi G. \and Ruiz-Lorenzo J.~J.}
  \REVIEW{Phys. Rev. B}{58}{1998}{2740}.


\end{thebibliography}
\end{document}